\documentclass[12pt,  onecolumn]{article}
\usepackage{amsmath}
\usepackage{amssymb}
\usepackage{graphicx}
\usepackage{caption2}
\usepackage{amsfonts}
\usepackage{cite}

\newcommand{\be}{\begin{equation}}
\newcommand{\ee}{\end{equation}}
\newcommand{\bea}{\begin{eqnarray}}
\newcommand{\eea}{\end{eqnarray}}
\newcommand{\bef}{\begin{figure}}
\newcommand{\ef}{\end{figure}}
\newcommand{\bt}{\begin{tabular}}
\newcommand{\et}{\end{tabular}}
\newcommand{\bno}{\begin{enumerate}}
\newcommand{\eno}{\end{enumerate}}

\def\3{\ss}

\catcode`\"=\active
\def"{\accent'177}

\begin{document}

\title{\huge Chaos: a bridge from microscopic  uncertainty\\
 to macroscopic randomness}

\author{Shijun Liao\\  \\ 
State Key Lab of Ocean Engineering,   Dept. of Mathematics\\
School of Naval Architecture, Ocean and Civil Engineering\\
 Shanghai Jiao Tong University,  Shanghai 200240, China\\ \\
( Email address:  sjliao@sjtu.edu.cn)}

\date{}

\maketitle

\hspace{-0.5cm}{\bf Abstract }   It is traditionally believed that the macroscopic randomness has nothing to do with the micro-level uncertainty.    Besides,  the sensitive dependence on initial condition (SDIC) of Lorenz chaos has never been considered {\em together} with the so-called continuum-assumption of fluid (on which Lorenz equations are based), from {\em physical} and {\em statistic} viewpoints.  A very fine numerical technique \cite{Liao2009} with negligible truncation and round-off errors,  called here the ``clean numerical simulation'' (CNS),  is applied  to investigate the propagation of the micro-level  unavoidable uncertain fluctuation (caused by the continuum-assumption of fluid) of initial conditions for  Lorenz equation with chaotic solutions.    Our  statistic analysis based on CNS computation of $10,000$ samples  shows that,  due to the SDIC,  the uncertainty of the micro-level statistic  fluctuation of initial conditions  transfers  into  the  macroscopic  randomness of  chaos.   This suggests that chaos  might  be  a  bridge  from  micro-level  uncertainty to macroscopic randomness, and thus would be an origin of macroscopic randomness.    We reveal in this article that, due to the SDIC of chaos and the inherent uncertainty of initial data, accurate long-term prediction of chaotic solution is not only {\em impossible} in mathematics but also has {\em no} physical meanings.  This  might provide us a new, different viewpoint to deepen  and enrich our understandings about the SDIC of chaos.     

\hspace{-0.5cm}{\bf Key Words} Chaos, propagation of uncertainty,  fine numerical simulation, multiple scales        

\section{Introduction: A paradox arising from Lorenz chaos}

Nowadays,  it is a common belief \cite{Egolf2000, Gaspard, Li1975, Smith1998,  Tucker1999,  Tsonis2008, Werndl2009}  of scientific society   that some  ``deterministic'' dynamic systems have chaotic behaviors: their solutions are exponentially sensitive to initial conditions so that accurate  long-term prediction of chaotic solution is  impossible.     Here, the deterministic means that the evolution of solutions is fully determined by  initial conditions {\em without}  random or uncertain  elements  involved.    Such kind of behaviors is called ``deterministic chaos'' \cite{Li1975},  because ``the deterministic nature of these systems does not make them predictable'' \cite{Werndl2009}.    

Such kind of non-periodic solutions  was first pointed out by Poincar\'{e}  \cite{Poincare1890} in 1880s for the famous three-body problem.    In 1962  Saltzman \cite{Saltzman1962}  found ``oscillatory, overstable cellular motions'' and ``consequently an alternating value of the heat  transport about a time-mean value'' for a free convection flow with very large Rayleigh number.   It is a pity that Saltzman \cite{Saltzman1962}  paid main attentions on the stable solutions for  Rayleigh number smaller than 10.   Fortunately,  this  ``oscillatory, overstable''  non-periodic solutions of the free convection flow  was  further  studied  in  details  by Lorenz \cite{Lorenz1963}  in 1963 for the weather prediction,  governed  by the so-called Lorenz equation 
\begin{eqnarray}
\dot{x} &=& \sigma \left( y - x \right), \\
\dot{y} &=& R \; x - y - x \; z, \\
\dot{z} & = & x \; y + b \; z,
\end{eqnarray}
where $\sigma, R$ and $b$ are physical parameters, the dot denotes the differentiation with respect to the time.  Although the Lorenz equation is much simpler than those used by Saltzman \cite{Saltzman1962},  its  solution  also becomes ``oscillatory, overstable'' for  large  Rayleigh  number.    Especially, using a digit computer and data in 6-digit precision,  Lorenz  \cite{Lorenz1963}  found that small changes in initial conditions leaded to great difference in long-term prediction, called today the ``butterfly effect''.   Based on the ``butterfly effect'',   Lorenz \cite{Lorenz1963}   made a correct conclusion  that  long-term weather prediction is impossible, although the Lorenz equation is only a very simple approximation model of the exact Navier-Stokes equations.   

All numerical methods have the so-called truncation and round-off error, more or less.  
Due to the so-called ``butterfly effect'',  {\em all} traditional numerical simulations of chaos    are  mixed  with such kind of  ``numerical noise''.   Unfortunately,   as pointed out by Lorenz \cite{Lorenz2006} in 2006, different traditional numerical schemes may lead to not only the uncertainty in prediction but also fundamentally different regimes  of the solution.   Thus, the traditional numerical simulations of chaos are not ``clean'' so that some of our understandings about  chaos based on these  impure numerical results  might be  questionable.        

 In order to gain reliable chaotic solutions in a long enough time interval,   Liao \cite{Liao2009}  developed a fine numerical technique with extremely high precision, called here the ``clean numerical simulation'' (CNS).   Using the computer algebra system Mathematica with the 400th-order Taylor expansion for continuous functions and data in accuracy of 800-digit precision,  Liao \cite{Liao2009}  gained, for the first time, ``clean'' numerical results of chaotic solution of Lorenz equation (in a special case $\sigma=10, R=28, b =-8/3$) 
 in a long time interval $0\leq t \leq 1000$  Lorenz time unit (LTU) with negligible truncation and round-off error.   It was found by Liao  \cite{Liao2009}  that,  to gain a reliable ``clean'' chaotic solution of Lorenz equation in $0\leq t \leq T_c$,  the initial conditions must be at least in the accuracy of  $10^{-2T_c/5}$.   Thus,  when $T_c = 1000$ LTU,  the initial condition must be  in the accuracy of 400-digit precision at least.    Currently, Liao's ``clean'' chaotic solution  \cite{Liao2009} of Lorenz equation is confirmed by Wang et al \cite{Wang2011}, who  used  parallel computation with the multiple precision (MP) library:   they gained reliable chaotic solution up to 2500 LTU by means of  the 1000th-order Taylor expansion and  data in 2100-digit precision, and their result agrees well  with Liao's one \cite{Liao2009} in $0\leq t \leq 1000$ LTU.    Their excellent work verified the validity of the ``clean numerical simulation'' (CNS) proposed by Liao \cite{Liao2009}.   These reliable ``clean'' chaotic solutions and especially the CNS provide us a powerful  tool to  investigate  the  essence  of SDIC and the ``butterfly effect'' from the {\em  physical} and {\em statistic} points of view, as shown below.    
 
Since Lorenz \cite{Lorenz1963} introduced the concept of SDIC of chaos,  its meanings has been discussed and investigated in many articles and books, mostly from the viewpoints of mathematics, logic and philosophy, but hardly from physical viewpoints.  This might be mainly because most models of chaos are too simple to accurately describe the complicated physical phenomena.  So, to deepen our understandings about the SDIC of chaos, it is valuable to study  it  from the physical viewpoints.            
           
Lorenz equation \cite{Lorenz1963}  was originally derived from the Navier-Stokes (N-S) equation describing phenomena of fluid motions.  The N-S equations  are based on such an  {\em assumption} that the fluid is a {\em continuum}, which is infinitely divisible and not composed of particles such as atoms and molecules.  Let us consider the uniform laminar flow of air with the velocity 1 (m/s) at the temperature $T = 0\; ^\circ$C  and the standard pressure.    In this case, there are  about $2.687 \times 10^{25}$  molecules in a cube of fluid.  This is a hugh number  so that  the continuum-assumption of fluid is mostly satisfied in practice.  Assume that {\em all} molecules of a cube of fluid have the same velocity, except {\em one}  which has a tiny  velocity fluctuation  $ 10^{-4}$ m/s.   Then, the averaged velocity fluctuation of a cube of fluid reads  $3.722 \times 10^{-30}$ m/s.    Such micro-level velocity fluctuation of fluid should be  neglected under the continuum-assumption.    
 In other words, in the frame of the continuum-assumption,  it has no {\em physical} meanings to consider the  {\em observable}  influence of such a tiny velocity fluctuation, from physical point of view!           

However,  Liao's CNS computation \cite{Liao2009}  in the accuracy of 800-digit precision  indicates that, {\em mathematically},  to gain reliable  chaotic solution in  $0 \leq T \leq 1000$ LTU, the fluctuation of  initial conditions must be less than 400-digit precision at least. Note that  the number $10^{-400}$  is  much  {\em smaller}  than $3.722 \times 10^{-30}$ that is a minimum of the averaged velocity fluctuation of fluid!   Thus,   as mentioned above,  from {\em physical} point of view,   such a tiny velocity fluctuation (in the level of $10^{-400}$)  has {\em no}  physical  meanings  at  all  under the continuum-assumption that is a base of Lorenz equation!     Therefore, a paradox  arises:  according to the continuum-assumption,  the tiny velocity fluctuation in the level of $10^{-30}$ should have no {\em observable}  influence on the chaotic solution of Lorenz equation;  on the other hand,  the SDIC and  ``butterfly effect''  indicate that the influence of a tiny velocity fluctuation even in the level of $10^{-400}$  must be considered!  This is certainly a paradox in logic!  

In history, many paradoxes first revealed the restrictions of some well-established theories and then greatly  promoted  their  developments.  What is the essence of this paradox from the viewpoint of physics?   What can we learn from it?

\section{From micro-level uncertainty to macroscopic randomness}

Without loss of generality, let us consider  the Lorenz equation   
with chaotic solution in case of  $R=28, b=-8/3$ and $\sigma=10$.  
Assume that the {\em observable}  values of initial condition \[ x_0 = -79/5, \;\;  y_0 = -437/25, \;\;  z_0 = 891/25\] are {\em given}  exactly.   However,  due to the continuum-assumption of fluid,  the initial conditions  involve the uncertainty:    the statistic fluctuations of velocity and temperature are {\em inherent} and {\em unavoidable} in essence, although their absolute values  are often much smaller than those of the observable values of initial condition.    According to the central limit theorem in probability theory,  we assume that the fluctuations of velocity and temperature are in the normal distribution with zero mean and a micro-level deviation $\sigma_0$, such as $\sigma_0 = 10^{-30}$ used in this article.    Thus, the {\em entire} initial conditions   
$ x(0)  =  x_0 +  \tilde x_{0},    y(0)  =  y_0 +  \tilde y_{0}$ and $z(0)  =  z_0 +  \tilde z_{0}$
involve  random,  where  $\tilde x_{0}, \tilde y_{0}, \tilde z_{0}$ are random variables in  the normal distribution with zero mean and deviation $\sigma_0$, i.e.
\[   \left<\tilde x_{0}\right> = \left<\tilde y_{0}\right> = \left<\tilde z_{0}\right>  = 0,   \hspace{0.5cm}
   \left<  \tilde x_{0}^2\right> = \left<  \tilde y_{0}^2\right> = \left<  \tilde z_{0}^2\right>= \sigma_0. \]
For each random initial condition, the corresponding ``clean''  chaotic solution is gained by means of the CNS \cite{Liao2009} with the 60-order Taylor expansion  and data in the accuracy of 120-digit precision.  For details, please refer to Liao \cite{Liao2009}.   According to Liao's work \cite{Liao2009},  both of the  truncation  and round-off error are negligible in $0\leq t < 180$ LTU.  Thus,  the  numerical results  are  ``clean'' at least in $0\leq t \leq 150$ LTU,  i.e.  without any {\em observable} influence by numerical noise.    Note that, although the standard deviation $\sigma_0 =10^{-30}$ of the uncertain terms $\tilde{x}_0, \tilde{y}_0, \tilde{z}_0$   of initial condition is much smaller than the observable values $x_0,y_0,z_0$, it is hugh compared to $10^{-120}$:  the truncation and round-off errors of the numerical simulations gained by the 60th-order Taylor formula and the data in accuracy of 120-digit precision are much smaller than the deviation $10^{-30}$ and thus are negligible in $0\leq t < 180$ LTU.   In this way, we can  accurately  investigate, {\em for the first time},   the   influence of the  micro-level statistic   fluctuation of initial conditions to chaotic solutions,  and especially the propagation of uncertainty from the micro-level  statistic fluctuation  of initial conditions  to  macroscopic randomness  of chaos.         

Let
$\left<x(t)\right>, \left<y(t)\right>, \left<z(t)\right>$ and  $\sigma_x(t), \sigma_y(t), \sigma_z(t)$  
denote the sample mean and  unbiased estimate of  standard deviation of $x(t), y(t), z(t)$, respectively, where $N=10^4$ is the number of samples gained by the CNS.    Define the so-called {\em uncertainty intensity} 
\begin{equation}
\epsilon(t)  =  \sqrt{\frac{[\sigma_x(t)]^2 +[\sigma_y(t)]^2+[\sigma_z(t)]^2}{\left<x(t)\right>^2+\left<y(t)\right>^2+\left<z(t)\right>^2}} \; . \label{def:epsilon}
\end{equation}
 
 \begin{figure}[thbp]
\centering
\includegraphics[scale=0.4]{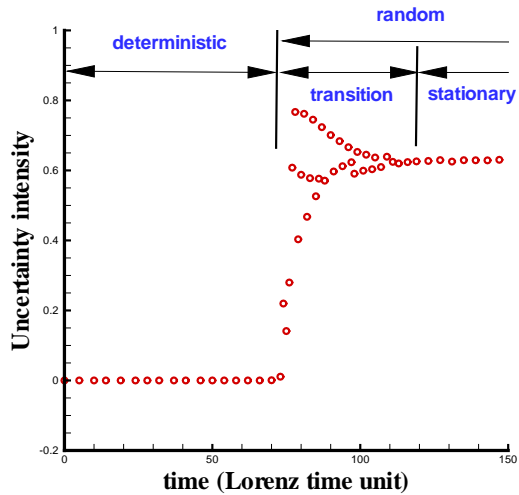}
\caption{The uncertainty intensity $\epsilon(t)$ in case of the fluctuation of initial conditions in the normal distribution with zero mean and micro-level deviation $\sigma_0 = 10^{-30}$.  }
\label{figure:epsilon}
\end{figure}

It is found that there exists such a time interval $t\in[0,T_d]$ with $T_d\approx 75$,  in which $\epsilon(t)$ is so small that one can  accurately predict the behavior of the dynamic system, but beyond which the uncertainty intensity increases greatly, as shown in Fig.~\ref{figure:epsilon}.   Thus,  using the result at any a point $t\in(0,T_d)$ as the initial condition and setting $t = -t$,   we  can  gain  the  given  {\em observable} values  $x_0,y_0,z_0$  of  initial conditions  in  a  high-level of accuracy,  meaning that the dynamic system looks like {\em deterministic} in $0\leq t \leq T_d$ and that  the influence of the uncertain  statistic fluctuation of initial condition is negligible.   But,  beyond it,  the solutions become rather sensitive to the uncertain statistic fluctuation (in the level of $10^{-30}$) of initial condition and  look like {\em random},   say,   the micro-level uncertain  statistic fluctuation in initial condition transfers into the observable macroscopic randomness.   So, $T_d$ is an important time scale for Lorenz chaos.             

As shown in Fig.~\ref{figure:epsilon},  there exists the time $T_s$ with $T_s\approx 120$ LTU, beyond which the cumulative distribution functions (CDF) of $x(t), y(t)$, $z(t)$ and so on are approximately stationary, i.e. almost independent of the time.   Besides,  these CDFs are independent of the observable values  $x_0,y_0,z_0$ of initial conditions,  meaning that all {\em observable} information of initial conditions are lost completely.  In other words,  when $t  > T_d$,  the asymmetry of time seems to break down so that the time has a one-way direction, i.e.  the arrow of time.    It suggests that, statistically, the chaotic Lorenz system might have two completely different dynamic behaviors before and after $T_d$:  it looks  like  ``deterministic'' without time's arrow when $t\leq T_d$, but thereafter rapidly becomes  random with the arrow of time.    This strongly suggests that chaos might be a bridge from the micro-level uncertainty to   macroscopic randomness, and  thus  might be an origin of macroscopic randomness and  the time's arrow.   This  provides us a new, different viewpoint to enrich and deepen our understandings about the SDIC of chaos.   

When $T_d < t < T_s$,  the CDFs of $x(t), y(t), z(t)$,  their sample means and unbiased estimates of  standard deviation  are time-dependent, and  evolve  to the approximately stationary ones  for $t\geq T_s$.   This process  is called  the {\em  transition} from the deterministic to randomness of chaos.   

\begin{figure}[thbp]
\centering
\includegraphics[scale=0.4]{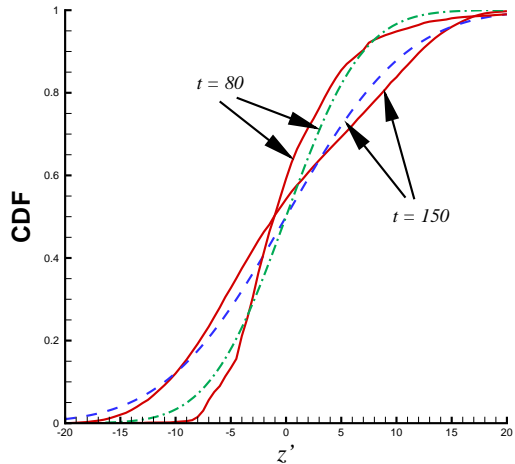}
\caption{The CDF (solid line) of $z'$ at $t = 80$ LTU and  $t = 150$ LTU, compared with the corresponding normal distribution (dashed line)  with the zero mean and the standard deviation $<z'^2>$.    }
\label{figure:CDF}
\end{figure}

Write $x' = x - <x>, y' = y-<y>$ and $z' = z-<z>$.    It is found that the CDFs of the fluctuations $x', y', z'$  are time-dependent when $t < T_s$ and become stationary when $t > T_s$.   When $t > T_d$,   the CDF  of  $x'$  is different from the normal distribution with the standard deviation $<x'^2>$, so are the CDFs of $y'$ and $z'$, as shown in Fig.~\ref{figure:CDF}.       It is also found that $T_d$ decreases exponentially with respect to $\sigma_0$,  the standard deviation of the tiny uncertain variables $\tilde{x}_0, \tilde{y}_0, \tilde{z}_0$  of the initial conditions.  Besides,   the stationary CDFs of $x',  y',  z'$ are independent of the CDFs of  $\tilde{x}_0, \tilde{y}_0, \tilde{z}_0$.   In addition, more samples are needed to gain accurate mean of the high correlations of $x', y', z'$, such as  $<x'z'>, <y' z'>$ and especially $<x'z'z'>, <x'y'y'>, <y'y'z'>,  <y'z'z'>$ and $<x'y'z'>$,  since the higher correlations  have  the larger standard derivations: this  shows the difficulty to propose an accurate model for the mean $<x>,  <y>,  <z>$ by means of  these higher correlations.   This also explains  why it is so difficult to propose a satisfied turbulence  model valid for all kinds of turbulent flows, since   Lorenz equation is a simplified model from Navier-Stokes equations.   Note that,   one can directly obtain all of these correlations from the Lorenz equation, as long as the number of  samples are large enough.   In other words,  no additional models for $<x>, <y>, <z>$ are needed.        

It is found that, given $\sigma_0=10^{-30}$ in case of $R=28, b=-8/3$ and $\sigma=10$,  we gain exactly the same figure as shown in Fig.~\ref{figure:epsilon}, even if we use more accurate numerical results obtained by means of the CNS with the 120-order Taylor expansion  and data in the accuracy of 240-digit precision!   Note that it has no physical meanings to use a  micro-level deviation $\sigma_0$ of the initial conditions smaller than $10^{-30}$,  as pointed out in the section of introduction.   Thus, for chaotic dynamic systems, the  transfer from micro-level uncertainty to macroscopic randomness seems unavoidable.    In addition,  it is fund that,  in case of  $b = -8/3, \sigma=10$ and  $R \leq  23.54$ so that solutions are not chaotic,  the micro-level uncertainty never transfers into the macroscopic level.  Therefore,  the SDIC of chaos is the key to such kind of transfer.   

\section{Conclusion and discussion}

In this article, the sensitive dependence on initial condition (SDIC) of Lorenz chaos is considered {\em together}, for the first time,  with the so-called continuum-assumption of fluid (on which Lorenz equations are based) from physical and  statistic viewpoints.   The so-called  ``clean numerical simulation'' (CNS) proposed by Liao  \cite{Liao2009}  is used to investigate the propagation of the micro-level  unavoidable uncertain  fluctuation (caused by the continuum-assumption of fluid) of initial conditions with chaotic solutions of Lorenz equation.    Our  statistic analysis based on the CNS computation of $10^4$ samples  suggests that,  due to the SDIC,  the uncertainty of the micro-level statistic  fluctuation of initial conditions  transfers  into  the  macroscopic  randomness of  chaos.    This may deepen and enrich  our understandings about the SDIC and chaos, from a different viewpoint of physics.         
                
The microscopic phenomena are  essentially uncertain, although probability distributions are governed by deterministic equations.   However,  it is traditionally believed that the micro-level uncertainty has no relationships with the macroscopic randomness.   But, our statistic analysis strongly suggests that  the micro-level uncertainty might be an origin of the macroscopic randomness, and chaos might be a bridge between them.   Although the above conclusion is based on Lorenz equation, it has general meanings.  First, we also investigated some other chaotic dynamic systems, and found the same transfer from micro-level uncertainty to macroscopic randomness for {\em all} of them.   Secondly,  as pointed out by Saltzman \cite{Saltzman1962}, the solutions of a  dynamic system consist of seven nonlinear differential equations for the free convention  (which is a more accurate model than Lorenz equation)  are  ``oscillatory, overstable''  (i.e. chaotic)  for large enough  Rayleigh  number.    In fact,  Saltzman \cite{Saltzman1962}  represented the solution of the original continuous differential equations as a sum of double-Fourier components,  and  approximated  the  original  problem  by a set of nonlinear ordinary different equations with finite number of degree of freedom.    Both of the Lorenz equation and the above mentioned system of seven equations are only special cases of it, corresponding to three and seven degree of freedom.   Obviously, the larger the degree of freedom, the more accurate the model.  It is found that, for large enough  Rayleigh  number,  these dynamic systems given by Saltzman \cite{Saltzman1962} with degree of freedom not less than three are chaotic, so that the micro-level uncertainty transfers into macroscopic randomness for all of them.  Theoretically speaking,   as  the  number  of  degree  of  freedom  tends to infinity,  this system becomes  the original continuous differential equations.  Thus,  our conclusion  about the transfer from micro-level uncertainty to macroscopic randomness has general meanings, although it is based on the Lorenz equation.   This  is  similar to the Lorenz's  famous conclusion  ``long-term accurate prediction of weather is impossible'' \cite{Lorenz1963},  which  is  based on the Lorenz equation, a very simple model of the N-S equation, but is correct and has been widely accepted by the scientific community.        

The similar transfer has been reported in  some other  fields.  For example,  as pointed out by Bai et al \cite{Bai},   the disorder of materials plays a fundamental role to the so-called sample-specific behavior of fracture, i.e.   the macroscopic failure may be quite different, sample to sample,  under the same macroscopic condition, because the differentiation due to meso-scopic disorder may be greatly amplified and lead to  largely different macroscopic effects.   Xia et al \cite{Xia2000}  studied the failure of disordered materials  by means of a stochastic slice sampling method with a nonlinear chain model, and found that   ``there is a sensitive zone in the vicinity of the boundary between the globally stable (GS) and evolution-induced catastrophic (EIC) regions in phase space, where a slight stochastic increment in damage can trigger a radical transition from GS to EIC''.   In other words, the meso-scopic uncertainty of disordered materials transfers into the macroscopic randomness of failure.  As mentioned by He et al \cite{He2004}, the nonlinearity and multi-scale might play a fundamental role in it.    So,   ``(stochastic) fluctuations are important and must not be neglected'' for the failure of disordered materials, as pointed out by Sahimi and Arbabi \cite{Sahimi1993}.   Another example is the evolution of the universe:  the micro-level uncertainty at Big Bang, the inherent  uncertainty of position and velocity of stars,  and the nonlinear property of gravity might be the origin of the macroscopic randomness of the universe.    All of these support our conclusion:  the transfer from micro-level  uncertainty to macroscopic randomness might have meanings in general.

Traditionally, it is believed that the SDIC of chaos  is  the origin of the so-called ``butter-fly effect'':  long-term prediction is impossible due to the SDIC of chaos and the impossibility of getting {\em exact} initial data with precision of arbitrary degree.  This traditional idea implies that the initial data itself {\em are} exact {\em inherently} but our human-being can not obtain the exact value.  However, as pointed out in this article, this traditional thought might be wrong:  due to the continuum-assumption of fluid,  there exists the statistic fluctuation of the initial data of Lorenz equation, {\em no matter} whether we could precisely measure the initial data or not.   It should be emphasized that such kind of uncertainty is {\em inherent}: it has nothing to do with our ability.    In this article, it is revealed that, due to the SDIC and the inherent uncertainty of initial data, accurate long-term prediction of chaotic solution is not only {\em impossible} in mathematics but also has {\em no} physical meanings.  This provides us a new explanation of the SDIC of chaos, from the physical and statistic points of view.           

The micro-level uncertainty and the physical variables $x, y, z$ of Lorenz equation are at different scales:  the absolute value of the former  (at the level of $10^{-30}$) is much smaller than that of the latter (at the level of 1).     Unfortunately,  the  truncation and round-off  errors  (often at the level of $10^{-10}$) of most traditional numerical techniques for chaos are much larger than such kind of micro-level uncertainty,  so that  the  propagation  of  the  micro-level  uncertainty  is  completely lost  in  the  numerical  noises.     The CNS \cite{Liao2009}  provides us a way to accurately investigate such kind of problems with multiple scales,  since the numerical noises of the CNS are much smaller than the micro-level uncertainty.         

 Lorenz equation is a simplified model based on the N-S equations describing flows of fluid.     
Note that nearly all models of turbulence are deterministic in essence: the micro-level uncertain statistic fluctuation of velocity caused by  the continuum-assumption of fluid has been neglected completely.  Note also  that the uncertainty intensity (\ref{def:epsilon}) is rather similar to the definition of turbulence intensity.   Since  turbulence  has a close relationship with chaos,  it might be possible that the  influence  of  the  micro-level  statistic  fluctuation of velocity and temperature   should be considered:  we even should  carefully  check   the  theoretical   foundation of  turbulence and the direct numerical simulation (DNS), such as the continuum-assumption of fluid.   Besides,   our very fine  numerical simulations and related analysis  reported in this article  suggest that the randomness of turbulence might come essentially from the micro-level uncertain statistic fluctuation of velocity and temperature: {\em turbulence is such a kind of flow of fluid that it is so unstable that the micro-level uncertainty transfers into macroscopic randomness}.    

Hopefully, this work stimulated by a  paradox  could provide us some new physical insights and mathematical ways to deepen and enrich our understanding about chaos and turbulence.   

\section*{Acknowledgement}  

 Thanks to the reviewers for their valuable comments and discussions.   The author would like to express his sincere thanks  to  Prof.  Y.L. Bai  and  Prof. M.F. Xia (Chinese Academy of  Sciences), Prof. Z. Li (Peking University), Prof. H.R. Ma (Shanghai Jiao Tong University) for their valuable discussions.   This work is partly supported  by State Key Lab of Ocean Engineering (Approval No. GKZD010053) and Natural Science Foundation of China (Approval No. 10872129).


\begin{thebibliography}{99.}

\bibitem{Bai}
Bai, Y. L., Ke, F.J., and  Xia, M. F.: Deterministically stochastic behavior and sensitivity to initial
configuration in damage fracture.  {\em Science Bulletin}  \textbf{39}:  892Ð895 (1994) [in Chinese].

 \bibitem{Egolf2000}
 Egolf,~D.A., Melnikov,~V.,  Pesch,~W. \&  Ecke,~R.E.:  Mechanisms of extensive  spatiotemporal chaos in
Rayleigh-B\`{e}nard convection. {\em Nature}, {\bf 404}: 733 -- 735 (2000).

 \bibitem{Gaspard}
Gaspard,~P.,  Briggs,~M.E.,   Francis,~M.K.,  Sengers,~J.V., 
Gammon, R.W., Dorfman, J.R. \& Calabrese, R.V.: Experimental evidence for microscopic chaos.  {\em Nature}, {\bf 394}: 865 -- 868  (1998).

\bibitem{He2004}
He, G.W., Xia, M.F., Ke, F.J., Bai, Y.L.: Multiple-scale coupled phenomena - challenge and opportunity.  {\em Progress in Natural Science}, \textbf{14}: 121 -- 124 (2004) [in Chinese].

\bibitem{Li1975}
 Li, T.Y. \& Yorke, J. A.:  Period three implies Chaos.  {\em American Mathematical Monthly},  \textbf{82}: 985 - 992 (1975).

\bibitem{Liao2009}
  Liao, S.J.:   On the reliability of computed chaotic solutions of non-linear differential equations.   {\em Tellus-A}, \textbf{61}: 550 -- 564  (2009).
  
  \bibitem{Lorenz1963}
Lorenz, E.N.:  Deterministic non-periodic flow.  {\em Journal of the Atmospheric Sciences}, \textbf{20}: 130 -- 141 (1963).

\bibitem{Lorenz2006} 
Lorenz,~E.N.: Computational periodicity as observed in a simple system.  {\em Tellus-A}, \textbf{58}: 549 -- 559 (2006). 

\bibitem{Poincare1890}
Poincar\'{e}, J.H.:  Sur le probl\`{e}me des trois corps et les \'{e}quations de la dynamique.  Divergence des s\'{e}ries de M. Lindstedt. {\em  Acta Mathematica}, \textbf{13}:1 -- 270 (1890).

 \bibitem{Sahimi1993}
  Sahimi, M. and Arbabi, S.: Mechanics of disordered solid. III. Fracture properties, {\em Physical Review B} \textbf{47}: 
 713 -- 722 (1993).
  
\bibitem{Saltzman1962}  
Saltzman,~B.: Finite amplitude free convection as an initial value problem (I).   {\em Journal of the Atmospheric Sciences}, \textbf{19}: 329 --  341, 1962. 
  

\bibitem{Smith1998}
Smith, P.: {\em Explaining Chaos}. Cambridge University Press, Cambridge (1998).   
   
\bibitem{Tsonis2008}
Tsonis,~A.A.:  {\em Randomnicity: Rules and Randomness in Realm of the Infinite}.  Imperial College Press (2008).  
  
  \bibitem{Tucker1999}   
Tucker,~W.:  The Lorenz  attractor exists.  {\em C. R. Acad. Sci.}  {\bf 328}:1197 -- 1202 (1999).
  
  \bibitem{Wang2011}
  Wang,~P.F.,  Li,~J.P. \& Li,~Q.:  Computational uncertainty and the application of a high-performance multiple
precision scheme to obtaining the correct reference solution of Lorenz equations.  {\em Numerical Algorithms},  online (DOI: 10.1007/s11075-011-9481-6).   
  

\bibitem{Werndl2009}
    Werndl, C.:  What are the new implications of Chaos for unpredictability?  {\em Brit. J. Phil. Sci.}  \textbf{60}:195 -- 220 (2009). 
    
\bibitem{Xia2000}
  Xia, M.F., Ke, F.J., Wei, Y.J., Bai, J. and Bai, Y.L.: Evolution induced catastrophe in a nonlinear dynamical model of material failure. {\em Nonlinear Dynamics} \textbf{22}: 205 -- 224 (2000).
  
    
\end{thebibliography}
\end{document}